\documentclass[useAMS,usenatbib]{mn2e}
\usepackage{graphicx,amssymb,amsmath,color,slashed,ulem,enumerate}

\title[DM--radiation interactions: DM haloes]
{Dark matter--radiation interactions: the impact on dark matter haloes}
\author[J.~A.~Schewtschenko et al.]
{J.~A.~Schewtschenko,$^{1,2}$\thanks{E-mail: j.a.schewtschenko@dur.ac.uk}
R.~J.~Wilkinson,$^2$
C.~M.~Baugh,$^1$
C.~B\oe hm,$^{2,3}$
S.~Pascoli$^2$\thanks{Also visiting Instituto de F\'{\i}sica Te\'orica, IFT-UAM/CSIC, Universidad Aut\'onoma de Madrid, Cantoblanco, 28049, Madrid, Spain}\vspace*{6pt}\\
$^1$Institute for Computational Cosmology, Durham University, Durham DH1 3LE, UK\\
$^2$Institute for Particle Physics Phenomenology, Durham University, Durham DH1 3LE, UK\\
$^3$LAPTH, U. de Savoie, CNRS,  BP 110, 74941 Annecy-Le-Vieux, France}

\date{\today}
\pubyear{2014}

\begin{document}

\label{firstpage}
\maketitle

%%%%%%%%%%%%%%%%%%%%%%%%%%%%%%%%%%%%%%%%%%%%%%%%%%%%%
%%%%%%%%%%%%%%%%%%%%%%%%%%%%%%%%%%%%%%%%%%%%%%%%%%%%%

\begin{abstract}
Interactions between dark matter (DM) and radiation (photons or neutrinos) in the early Universe suppress density fluctuations on small mass scales. Here we perform a thorough analysis of structure formation in the fully non-linear regime using $N$-body simulations for models with DM--radiation interactions and compare the results to a traditional calculation in which DM only interacts gravitationally. Significant differences arise due to the presence of interactions, in terms of the number of low-mass DM haloes and their properties, such as their spin and density profile. These differences are clearly seen even for haloes more massive than the scale on which density fluctuations are suppressed. We also show that semi-analytical descriptions of the matter distribution in the non-linear regime fail to reproduce our numerical results, emphasizing the challenge of predicting structure formation in models with physics beyond collisionless DM.
\end{abstract}


%%%%%%%%%%%%%%%%%%%%%%%%%%%%%%%%%%%%%%%%%%%%%%%%%%%%%
\section{Introduction}
\label{sec:intro}
%%%%%%%%%%%%%%%%%%%%%%%%%%%%%%%%%%%%%%%%%%%%%%%%%%%%%

Dark matter (DM) is the most dominant and yet most elusive component of matter in the Universe. Exploring its nature is therefore one of the greatest challenges in both cosmology and particle physics today. The usual treatment of DM in structure formation calculations neglects possible interactions between DM and other species. Yet if DM is a (thermal) weakly interacting massive particle (WIMP), interactions (and more precisely, annihilations) are essential to obtain the correct relic density. It is therefore important to study the impact of DM interactions on other cosmological observables. 

It has been already established that a DM coupling with primordial radiation, i.e. photons~(\citealt{Boehm:2000gq,boehm_interacting_2001,Sigurdson:2004zp,Boehm:2004th,Dolgov:2013una,Wilkinson:2013kia}) or neutrinos~(\citealt{Boehm:2000gq,boehm_interacting_2001,Boehm:2004th,Mangano:2006mp,Serra:2009uu,Wilkinson:2014ksa}) leaves a characteristic imprint on the CMB temperature and polarization power spectra.
In addition, in a previous publication~(\citealt{boehm:2014MNRAS}), we showed using $N$-body simulations that such interactions have a significant impact on the Milky Way environment, dramatically reducing the number of DM subhaloes that could potentially host satellite galaxies\footnote{See also \cite{Bertoni:2014mva}.}. Since they have the potential to alleviate the small-scale problems that have persisted in the standard cold DM (CDM) model for more than a decade~(\citealt{moore_dark_1999,Klypin:1999uc,BoylanKolchin:2011de}), these interactions should not be ignored. 

We now go a step further and study the abundance and properties, such as shape, spin and density profile of collapsed DM structures in the presence of DM--radiation interactions. We highlight the differences with respect to CDM and in addition, warm DM (WDM), which shows a qualitatively similar suppression of power on small scales~(\citealt{schaeffer_silk}). We note that recent work has also considered non-linear structure formation in a number of alternative models such as self-interacting DM (\citealt{Rocha:2012jg,Vogelsberger:2014pda}), decaying DM (\citealt{Wang:2014ina}), late-forming DM (\citealt{Agarwal:2014qca}), atomic DM (\citealt{CyrRacine:2012fz}) and DM interacting with dark radiation (\citealt{Buckley:2014PhRvD,Chu:2014lja}); see also \cite{Schneider:2014rda}.

The paper is organized as follows. In Sec.~\ref{sec:theory}, we summarise the theoretical background and results obtained thus far using linear perturbation theory. In Sec.~\ref{sec:simulations}, we describe the setup of our numerical simulations. In Secs.~\ref{sec:hmf} and~\ref{sec:spinconc}, we analyse the abundance and properties of the collapsed structures, comparing our results with semi-analytical approximations from the literature. Our conclusions are presented in Sec.~\ref{sec:conc}.

%%%%%%%%%%%%%%%%%%%%%%%%%%%%%%%%%%%%%%%%%%%%%%%%%%%%%
\section{Theoretical Background}
\label{sec:theory}
%%%%%%%%%%%%%%%%%%%%%%%%%%%%%%%%%%%%%%%%%%%%%%%%%%%%%

Among all the possible contributions to the collisional damping of DM fluctuations, the largest occurs when DM interacts with photons ($\gamma$CDM) or neutrinos ($\nu$CDM). There are two reasons for this: (i) photons and neutrinos have the largest energy density of any standard model particle until matter-radiation equality, (ii) they are relativistic and therefore tend to drag DM particles out of small mass overdensities if they are coupled to DM.

For large values of the DM--radiation scattering cross-section, the suppression is prominent in both the CMB temperature and polarization power spectra. A comparison between the predicted spectra and the first-year data from {\it Planck}~(\citealt{Ade:2013zuv}) gives upper bounds of $8 \times 10^{-31}~(m_{\rm DM}/\mathrm{\mathrm{GeV}})~ \mathrm{cm}^2$ and $2 \times 10^{-28}~(m_{\rm DM}/\mathrm{\mathrm{GeV}})~ \mathrm{cm}^2$ on the $\gamma$CDM and $\nu$CDM cross-sections respectively, where $m_{\rm DM}$ is the DM particle mass (at 68\% CL, assuming a constant cross-section)~(\citealt{Wilkinson:2013kia,Wilkinson:2014ksa}).

\begin{figure}
\includegraphics[width=0.49\textwidth]{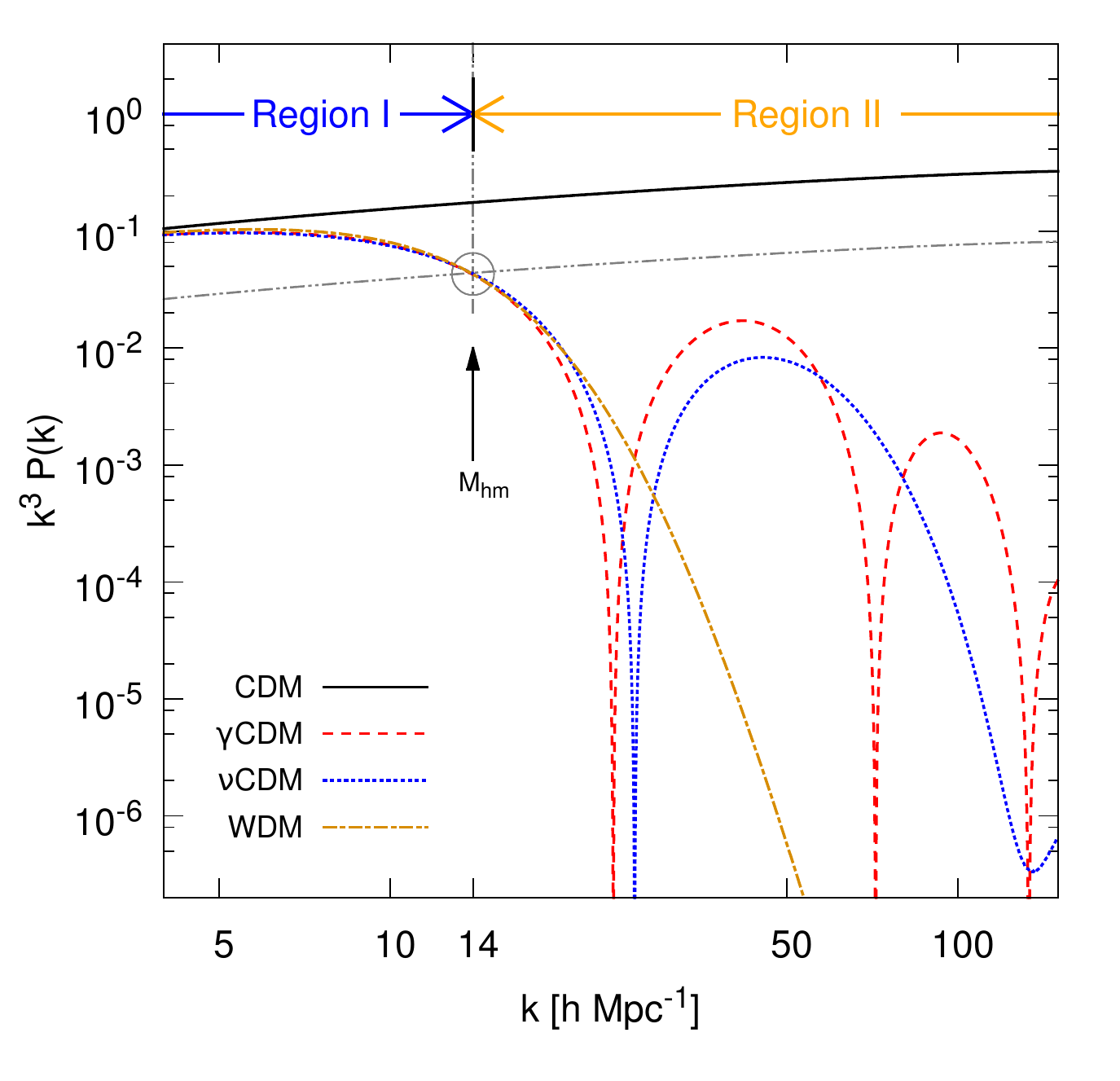}
\vspace{-5ex}
\caption{The linear matter power spectra for collisionless CDM (solid, black), $\gamma$CDM (dashed, red), $\nu$CDM (dotted, blue) and WDM (dashed-dotted, orange) at redshift $z = 49$. The interaction cross-sections for $\gamma$CDM and $\nu$CDM and the particle mass for WDM have been selected such that the initial suppression with respect to CDM is identical (see Table~\ref{tab:params}). This wavenumber defines the {\it half-mode} mass, $M_{\rm hm}$, which is marked with an arrow and separates regions I and II, which are discussed with reference to Fig.~\ref{fig:theory:damp_scale}.}
\label{fig:theory:ps}
\end{figure}

The reason why these constraints differ for $\gamma$CDM and $\nu$CDM is that photons and neutrinos do not have exactly the same effect on DM fluctuations due to their different thermal histories, with photons staying coupled to the thermal bath for much longer due to Thomson scattering\footnote{In addition, $\gamma$CDM has a direct impact on the CMB, while $\nu$CDM only affects the CMB indirectly, and the parameter space for $\nu$CDM suffers from significant degeneracies~(see~\citealt{Wilkinson:2014ksa}).}. Their effect on the matter power spectrum is also different, as illustrated in Fig.~\ref{fig:theory:ps}, where we show the linear theory matter power spectra for collisionless CDM, $\gamma$CDM, $\nu$CDM and (collisionless) WDM.

Unless explicitly stated otherwise, the values we use throughout this paper for the $\gamma$CDM and $\nu$CDM cross-sections and the WDM mass are given in Table~\ref{tab:params}. These parameters are motivated by the constraints obtained in our previous work~(\citealt{boehm:2014MNRAS}) and have been selected such that the scale at which the transfer function is suppressed by a factor of two with respect to CDM (hence giving a factor of four reduction in power) is identical. This scale defines the {\it half-mode} mass, $M_{\rm hm}$, and demarks the range of wavenumbers labelled as regions I and II in Fig.~\ref{fig:theory:ps}. In region II, there are important differences between the power spectra for $\gamma$CDM, $\nu$CDM and WDM.

\begin{table}
\begin{center}
  \begin{tabular}{c|ccc}
     \hline\hline
~& $(m_{\rm DM}$/GeV) & $(m_{\rm DM}$/GeV) & $(m_{\rm DM}/{\rm g})$ \\
~& $\times~\sigma_{\rm Th}$ & $\times~{\rm cm}^2$ & $\times~{\rm cm}^2$ \\ \hline 
$\gamma$CDM & $2.0 \times 10^{-9}$ & $1.3 \times 10^{-33}$ &
     $7.5 \times 10^{-10}$ \\
$\nu$CDM & $2.9 \times 10^{-9}$ & $1.9 \times 10^{-33}$ &
     $1.1 \times 10^{-9}$ \\ \hline\hline
   \end{tabular}
   \vspace{3ex}
  \begin{tabular}{c|cc}
     \hline\hline
& $m_{\rm DM}$~[keV] & $\alpha$~[${h}^{-1}~{\rm Mpc}$] \\ \hline 
WDM & 1.2 & 0.037 \\ \hline\hline
   \end{tabular}
   \vspace{2ex}
\caption{The (constant) elastic scattering cross-sections for $\gamma$CDM and $\nu$CDM and the particle mass for WDM, expressed in various units. $\sigma_{\rm Th}$ is the Thomson cross-section, $m_{\rm DM}$ is the DM mass and $\alpha$ is defined in Eq.~\eqref{eq:alpha}. Note that the mass of the WDM particle is motivated by the reproduction of the half-mode mass in the $\gamma$CDM and $\nu$CDM models, as explained in the text.}
\label{tab:params}
\end{center}
\end{table}

In the case of a thermalized, non-interacting, fermionic WDM particle, the suppression in the matter power spectrum is typically approximated by the transfer function~(\citealt{Bode:2001ApJ})
\begin{equation}
T(k) = \left[1+(\alpha k)^{2 \mu}\right]^{- 5/ \mu} \ ,
\end{equation}
where
\begin{equation}
 \alpha = 0.048 \left[\frac {m_{\rm DM}} {\rm keV} \right]^{-1.15} \left[\frac{\Omega_{\rm DM}} {0.4} \right]^{0.15} \left[ \frac h {0.65} \right]^{1.3} \frac {\mathrm{Mpc}} h \ .
 \label{eq:alpha}
\end{equation}
Here, $\Omega_{\rm DM}$ is the DM energy density, $h$ is the reduced Hubble parameter and $\mu \simeq 1.2$ is a fitting parameter\footnote{There is an alternative fit for $\alpha$ and $\mu$ that is often used in the literature~(e.g. \citealt{Viel:2005PhRvD}), but the difference is marginal for our analysis.}. The scale $\alpha$ in Eq.~\ref{eq:alpha} encapsulates the effect of free-streaming, which erases primordial fluctuations below a wavelength given by
\begin{equation}
 \lambda_{\rm fs} = \int_{t_{\rm dec}}^{t_0} \frac{v(t)} {a(t)} \mathrm{d}t~\approx~r_{\rm H}(t_{\rm NR}) \left[1+\frac 12 \log \left(\frac {t_{\rm EQ}}{t_{\rm NR}} \right)\right]~,
\label{eq:theory:fs}
\end{equation}
where $v(t)$ is the thermal velocity of the WDM particle. In this expression, $t_{\rm dec}$ is the DM decoupling time, $t_0$ is the time today, $a(t)$ is the cosmological scale factor, $r_{\rm H}(t_{\rm NR})$ is the comoving size of the horizon when DM becomes non-relativistic (at time $t_{\rm NR}$) and $t_{\rm EQ}$ is the epoch of matter-radiation equality.

A similar transfer function can be used to model the cut-off in the matter power spectra in $\gamma$CDM and $\nu$CDM~(\citealt{boehm_interacting_2001}) with
\begin{equation}
 \tilde \alpha = \beta_X \left[ \frac {\sigma_{{\rm DM-}X}} {\sigma_{\rm Th}} \frac {m_{\rm DM}} {\rm \mathrm{GeV}} \right]^{0.48} \left[\frac{\Omega_{\rm DM}} {0.4} \right]^{0.15} \left[ \frac h {0.65} \right]^{1.3} \frac {\mathrm{Mpc}} h \ ,
 \label{eq:tkint}
\end{equation}
where $X$ is $\gamma$ or $\nu$, $\beta_\gamma \approx 1.25\times10^4$, $\beta_\nu \approx 1.04\times10^4$, ${\sigma_{{\rm DM-}X}}$ is the DM--radiation cross-section and $\sigma_{\rm Th}$ is the Thomson cross-section. This transfer function fixes the half-mode scale for $\gamma$CDM and $\nu$CDM, thus providing a means to compare the impact of the interactions with respect to WDM, but does not encapsulate the full suppression of the power spectrum.

Eq.~\eqref{eq:tkint} corresponds to an analytical calculation of the collisional damping scale given by\footnote{We neglect the possible contributions from self-interactions and {\it mixed} damping and simplify the calculation to a single DM interaction partner.}
\begin{equation}
 \lambda^2_{\rm cd} = \frac {2 \pi^2} 3 \int_0^{t_{\rm dec}} \frac {\rho_X} {\slashed\rho} \frac {v_X^2~(1+\Theta_X) }{a^2~\Gamma_X}~\mathrm{d}t~.
\label{eq:theory:cd1}
\end{equation}
In this equation, $\slashed \rho = \rho_X + p_X$, where $\rho_X$ is the energy density, $p_X$ the pressure, $v_X$ is the velocity dispersion and $\Gamma_X$ is the total interaction rate of the DM interaction partner and $\Theta_X$ contains the contribution from heat conduction.

As the integral in Eq.~\eqref{eq:theory:cd1} is dominated by the contribution at late times, the collisional damping scale can be approximated by
\begin{equation}
\lambda^2_{\rm cd} \approx \frac {2 \pi^2} 3~\left[ \frac {\rho_X} {\slashed\rho} \ \frac {v_X^2~(1+\Theta_X)}{a^2}~\frac{t^2}{\alpha_X}\right]~\bigg|_{t_{\rm dec}}~,
\label{eq:theory:cd}
\end{equation}
using $\Gamma_X = H = \alpha_X/t$ at $t = t_{\rm dec}$, where $H$ is the Hubble rate, $\alpha_X = 1/2$ if $t_{\rm{dec}} < t_{\rm EQ}$ and $\alpha_X = 2/3$ otherwise. On scales smaller than $\lambda_{\rm cd}$, primordial fluctuations are erased.

\begin{figure}
\includegraphics[width=0.46\textwidth]{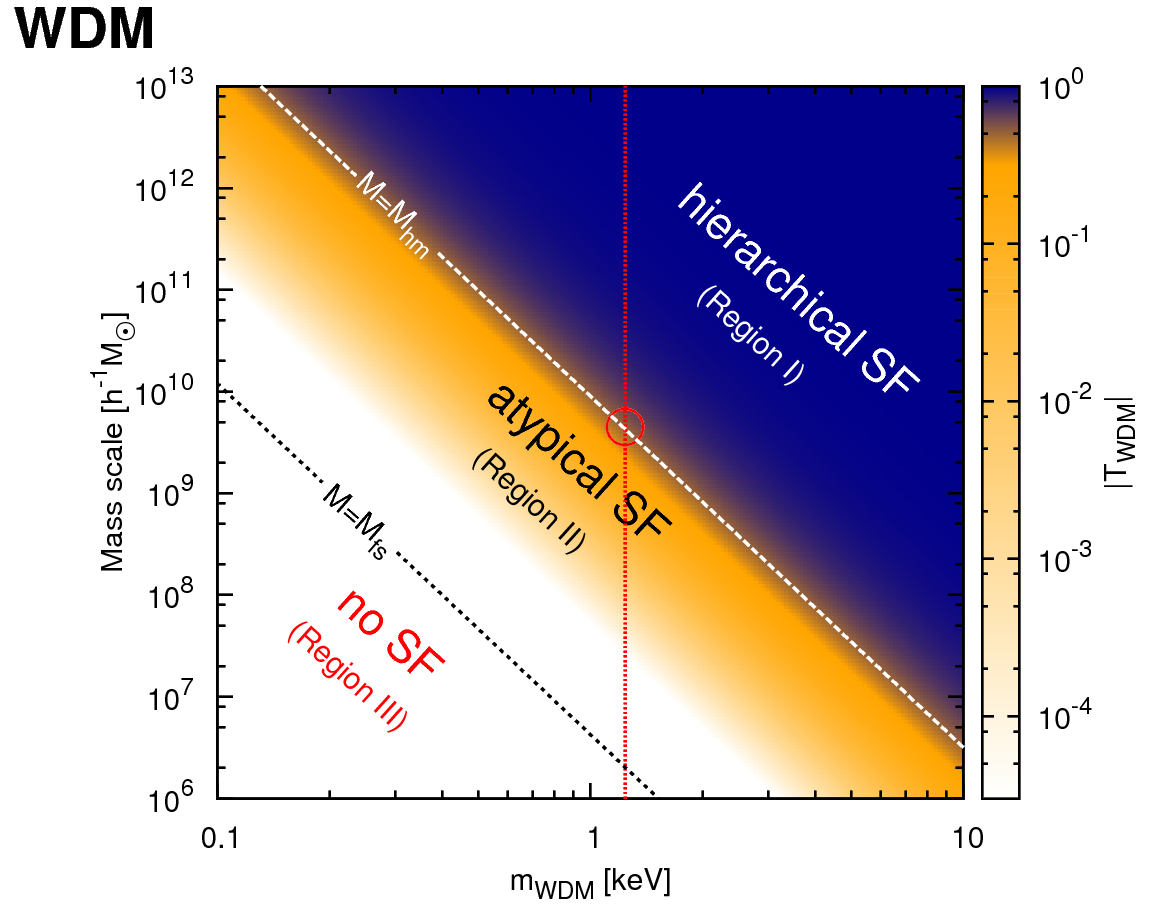}
\includegraphics[width=0.46\textwidth]{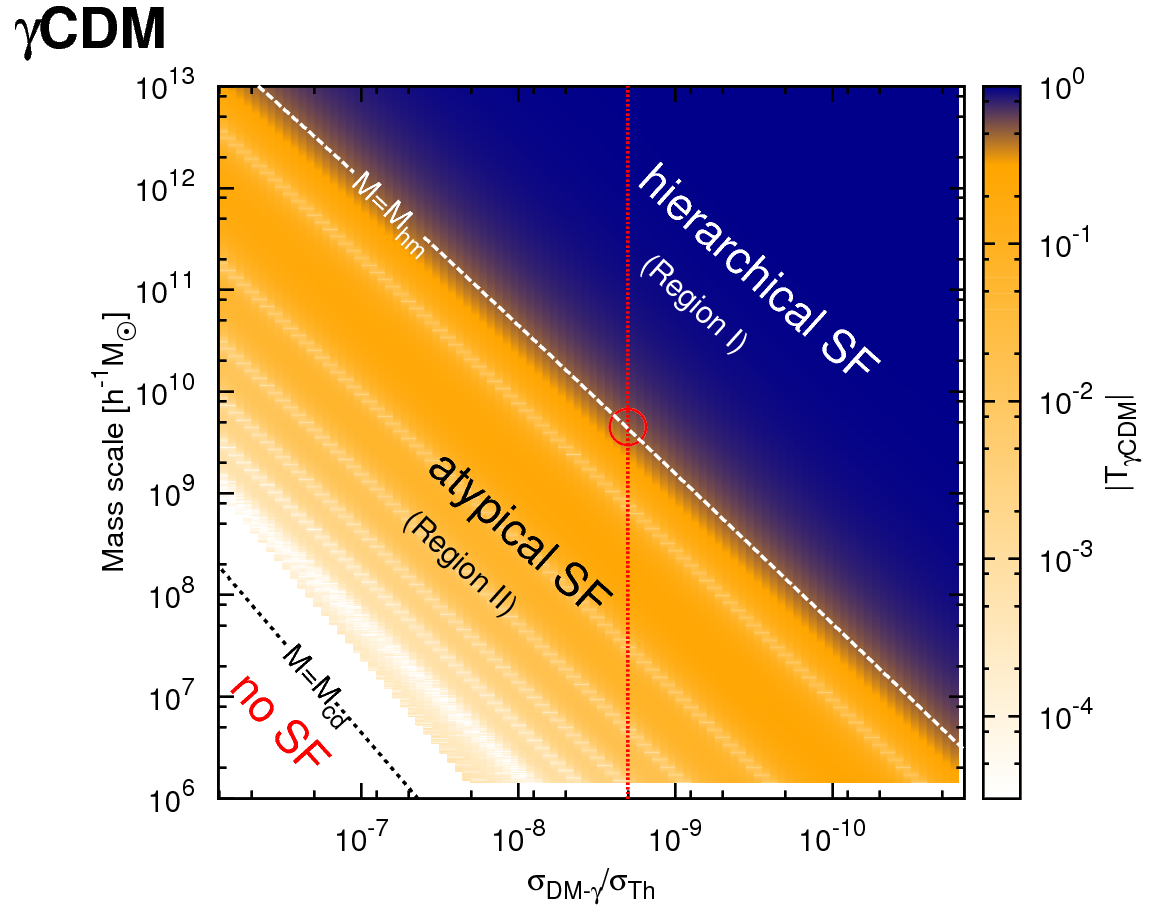}
\includegraphics[width=0.46\textwidth]{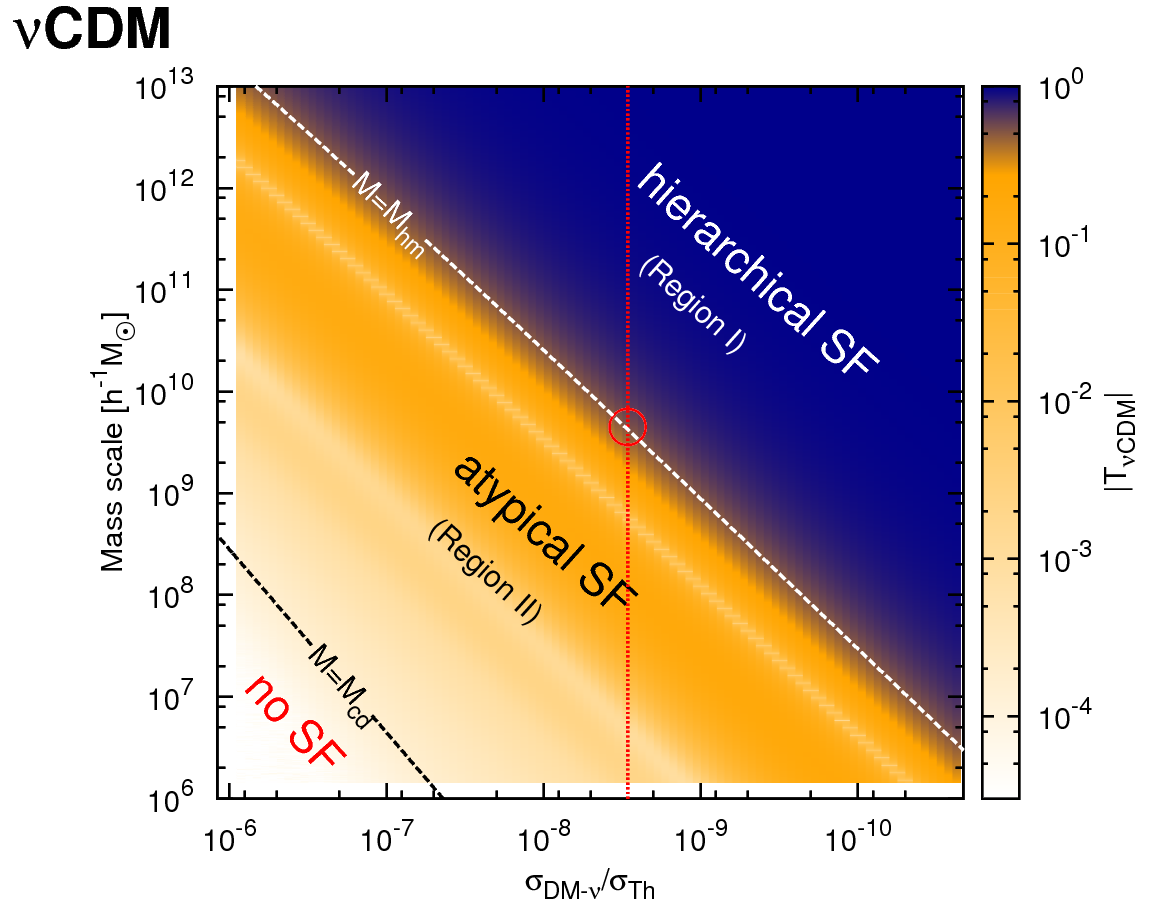} 
\caption{Characteristic mass scales for the suppression of primordial fluctuations by free-streaming (WDM, top), photon collisional damping ($\gamma$CDM, middle) and neutrino collisional damping ($\nu$CDM, bottom). The half-mode mass scale, $M_{\rm hm}$, is defined by the initial cut-off in the transfer function and marks the upper boundary of region II, where hierarchical structure formation (SF) may no longer occur due to a reduced number of low-mass progenitors. $M_{\rm fs}$ and $M_{\rm cd}$ are the masses corresponding to the free-streaming and collisional damping scales respectively and define the boundary of region III, where structures no longer form. The colour scale shows the absolute value of the transfer function, $T(k)$, and the vertical red lines correspond to the DM parameters listed in Table~\ref{tab:params}.}
\label{fig:theory:damp_scale}
\end{figure}

We summarise the impact of the damping scales $\lambda_{\rm fs}$ and $\lambda_{\rm cd}$ in linear theory in Fig.~\ref{fig:theory:damp_scale}. To distinguish these quantities from the half-mode mass scale, $M_{\rm hm}$, we present the mass corresponding to the relevant damping scale as a function of the DM mass (for WDM) and interaction cross-section (for $\gamma$CDM and $\nu$CDM).

We identify three regions in Fig.~\ref{fig:theory:damp_scale}. Regions I and II are already labelled in Fig.~\ref{fig:theory:ps} and there is now an additional region (III) occurring at much higher wavenumbers than are plotted in this figure. In region I, haloes form hierarchically, while in region III, all primordial perturbations have been erased. In between lies a transition region (region II), where some primordial density fluctuations may survive to form structure, but these are already sufficiently suppressed to disfavour a typical hierarchical structure formation. Region II extends down to much smaller scales for $\gamma$CDM and $\nu$CDM compared to WDM due to the prominent oscillations in the matter power spectrum\footnote{We note that acoustic oscillations are also expected in the transfer functions for certain WDM models at small scales (see e.g. \citealt{Boyanovsky:2010pw}). However, at these scales, the transfer function is already strongly suppressed by free-streaming so the regeneration of power from these oscillations is expected to be much weaker than in $\gamma$CDM and $\nu$CDM.}. The separation between regions I and II is determined by the half-mode mass scale (as in Fig.~\ref{fig:theory:ps}), while the transition between regions II and III is governed by the free-streaming scale (for WDM) or collisional damping scale (for $\gamma$CDM and $\nu$CDM).

%%%%%%%%%%%%%%%%%%%%%%%%%%%%%%%%%%%%%%%%%%%%%%%%%%%%%
\section{Simulations}
\label{sec:simulations}
%%%%%%%%%%%%%%%%%%%%%%%%%%%%%%%%%%%%%%%%%%%%%%%%%%%%%

To calculate the non-linear evolution of the matter distribution, we run a suite of high-resolution $N$-body simulations using the parallel Tree-Particle Mesh code, \texttt{GADGET-3}~(\citealt{gadget2}). To model a wide dynamical range, we perform simulations in large boxes (of side lengths 100 h${}^{-1}$~Mpc and 300 $h^{-1}$~Mpc) and a small box (of side length 30 $h^{-1}$~Mpc), all containing $1024^3$ particles.

The simulations begin at a redshift of $z = 49$ (the DM--radiation interaction rate is negligible for $z < 49$) and use a gravitational softening of 5\% of the mean particle separation. The initial conditions are created with an adapted version of a second-order LPT code~(\citealt{2lptic_refs}), using input matter power spectra from a modified version of the Boltzmann code, \texttt{CLASS}~(\citealt{class_refs}). 

We use the best-fitting values of the cosmological parameters obtained by the {\it Planck} collaboration in the ``{\it Planck} + WP'' dataset~(\citealt{Ade:2013zuv}), assuming a flat $\Lambda$CDM cosmology. In principle, a consistent treatment of an interacting DM model would require one to study each cross-section within its own best-fitting cosmology. However, we find that the parameters for $\Lambda$CDM lie within one standard deviation of such best fits. Therefore, we keep the cosmological parameters fixed for all the models studied here.

\cite{Lovell:2013ola} showed that in the case of WDM, one can safely ignore thermal velocities, without introducing a significant error on the scales of interest, if the DM particle is heavier than $\sim$ 1 keV. We confirmed this by performing simulations with and without a thermal velocity dispersion and obtaining convergence on the scales of interest. Hence, we only consider models in which late-time free-streaming can be neglected.

\begin{figure*}
\includegraphics[width=0.47\textwidth]{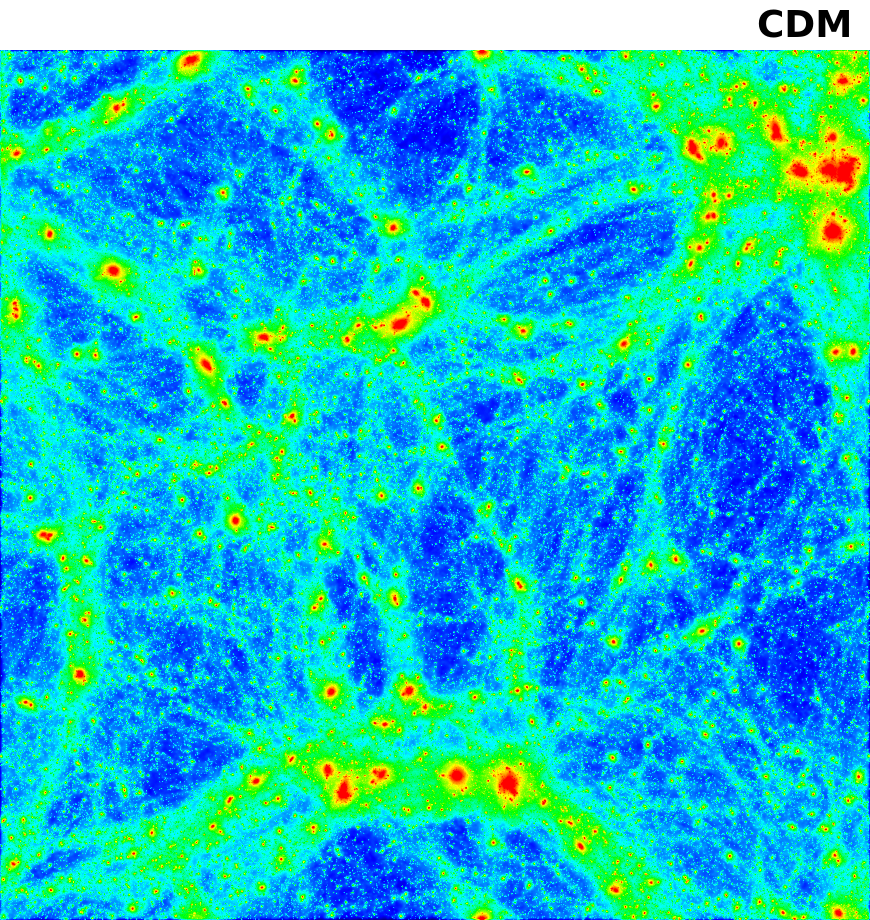}
\includegraphics[width=0.47\textwidth]{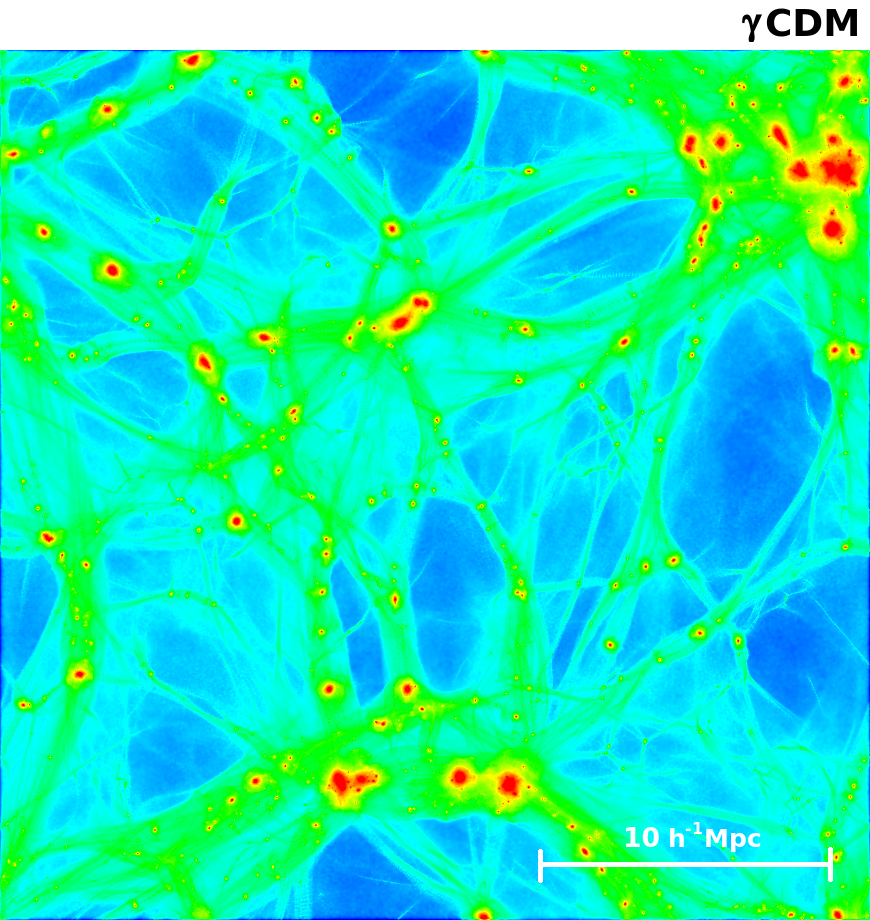}
\vspace{2ex}
\caption{The simulated distribution of DM at redshift $z = 0$ in a box of side length 30 $h^{-1}$~Mpc for two models: collisionless CDM (left) and $\gamma$CDM with $\sigma_{{\rm DM}-\gamma} = 10^{-7} \hspace{0.5ex} \sigma_{\rm Th} \hspace{0.5ex} (m_{\rm DM}/\mathrm{\mathrm{GeV}})$ (right), which is allowed by {\it Planck} CMB data~(\citealt{Wilkinson:2013kia}). The colours indicate the DM density on a scale increasing from blue to red. Due to collisional damping, we obtain fewer small-scale structures in $\gamma$CDM than are seen in CDM.}
\label{fig:sim:projection}
\end{figure*}

Fig.~\ref{fig:sim:projection} shows the projected DM distribution in the 30 $h^{-1}$~Mpc box for (i) collisionless CDM and (ii) an extreme $\gamma$CDM model that is allowed by {\it Planck} CMB data (\citealt{Wilkinson:2013kia}). Fewer small structures are present in $\gamma$CDM as an immediate result of the suppression of small-scale power shown in Fig.~\ref{fig:theory:ps}. The only exception is found along the filaments, where spurious structures contaminate the otherwise smooth environment (\citealt{Wang:2007MNRAS}). Similar results are obtained for $\nu$CDM and WDM.

For abundance measurements (Sec.~\ref{sec:hmf}), DM haloes are identified using a {\it friends-of-friends} group finder (\citealt{Davis:1985rj}) with a linking length of 20\% of the mean particle separation. For the halo properties (Sec.~\ref{sec:spinconc}), we instead use the {\it AMIGA halo finder}~(\citealt{ahf_refs}), where collapsed structures are defined as spherically overdense regions of radius $r_{\rm vir}$ with a mean density given by
\begin{equation}
\frac {3 M_{\rm vir}}{4 \pi r^3_{\rm vir}} = \Delta_{\rm th} \rho_{\rm crit} \label{eq:sim:halodef}~.
\end{equation}
In this expression, $M_{\rm vir}$ is the virial mass, $\rho_{\rm crit}$ is the critical density and $\Delta_{\rm th}$ is the mean overdensity of a virialized halo with respect to the critical density, according to the spherical top-hat collapse model.

%%%%%%%%%%%%%%%%%%%%%%%%%%%%%%%%%%%%%%%%%%%%%%%%%%%%%
%%%%%%%%%%%%%%%%%%%%%%%%%%%%%%%%%%%%%%%%%%%%%%%%%%%%%
\section{Results: Halo Abundance}
\label{sec:hmf}
%%%%%%%%%%%%%%%%%%%%%%%%%%%%%%%%%%%%%%%%%%%%%%%%%%%%%
%%%%%%%%%%%%%%%%%%%%%%%%%%%%%%%%%%%%%%%%%%%%%%%%%%%%%

The suppression of small-scale density fluctuations in the early Universe (as discussed in Sec.~\ref{sec:theory}) has a significant effect on the subsequent structure formation. This has been studied in detail for WDM (e.g. \citealt{Lovell:2013ola}), where the halo mass function (HMF) was compared to semi-analytical predictions. In this section, we perform a similar analysis for $\gamma$CDM and $\nu$CDM by comparing the simulated HMFs with the {\it Press-Schechter} formalism (\citealt{Press:1974ApJ}) and modifications thereof. In addition, we study the spatial distribution of DM haloes on large scales. 

%%%%%%%%%%%%%%%%%%%%%%%%%%%%%%%%%%%%%%%%%%%%%%%%%%%%%
\subsection{Semi-Analytical Halo Mass Functions}
\label{subsec:psf}
%%%%%%%%%%%%%%%%%%%%%%%%%%%%%%%%%%%%%%%%%%%%%%%%%%%%%

The {\it Press-Schechter} formalism uses the known primordial perturbations and their linear growth to calculate the fractional volume of space occupied by virialized objects of a given mass, assuming a spherical collapse model~(\citealt{Press:1974ApJ}). The halo mass function (HMF) can be written as
\begin{equation}
 \frac {\mathrm{d}n(M)}{\mathrm{d}M} = - \frac{1}{2}~f^{\rm (HMF)}(\sigma^2)~\frac{\bar \rho}{M^2} \frac {{\rm d} \ln \sigma^2(M)} {{\rm d} \ln M}\ ,
\label{eq:frac}
\end{equation}
where $n(M)$ is the number density of DM haloes of mass $M \rightarrow M+{\rm d}M$, $\bar \rho$ is the average matter density of the Universe and $\sigma^2(M)$ is the variance of the linear density field given by
\begin{equation}
\sigma^2(M) = \frac 1 {2 \pi^2} \int_0^\infty k^2 P(k) \hat{W}^2(k,R)~{\rm d}k~.
\end{equation}
The variance is smoothed on a mass-dependent scale $R(M)$, using a suitable window function $W(r,R)$, which has a Fourier transform $\hat{W}(k,R)$~(\citealt{Jenkins:2001MNRAS}).

The {\it Sheth-Tormen} (ST) formalism~(\citealt{Sheth:2001MNRAS}) combines the Press-Schechter formalism with an ellipsoidal collapse model. In this model, the function $f^{\rm (HMF)}(\sigma^2)$ in Eq.~\ref{eq:frac} represents the fraction of collapsed haloes and is defined by
\begin{equation}
 f^{\rm (HMF)}_{\rm ST}(\sigma^2) = A \sqrt{\frac {2} \pi} \left[ 1 + x^{-2p} \right] x \exp\left[ - x^{2}/2 \right]~.
\end{equation}
In this expression, $x \equiv \sqrt{a} \delta_c/\sigma$, where $\delta_c$ is the cosmology-dependent linear overdensity at the time of collapse. The parameters $A \approx 0.3222$, $p \approx 0.3$ and $a \approx 0.707$ were obtained by fitting to simulation results \citep{Sheth:2001MNRAS}.

\begin{figure}
\includegraphics[width=0.55\textwidth]{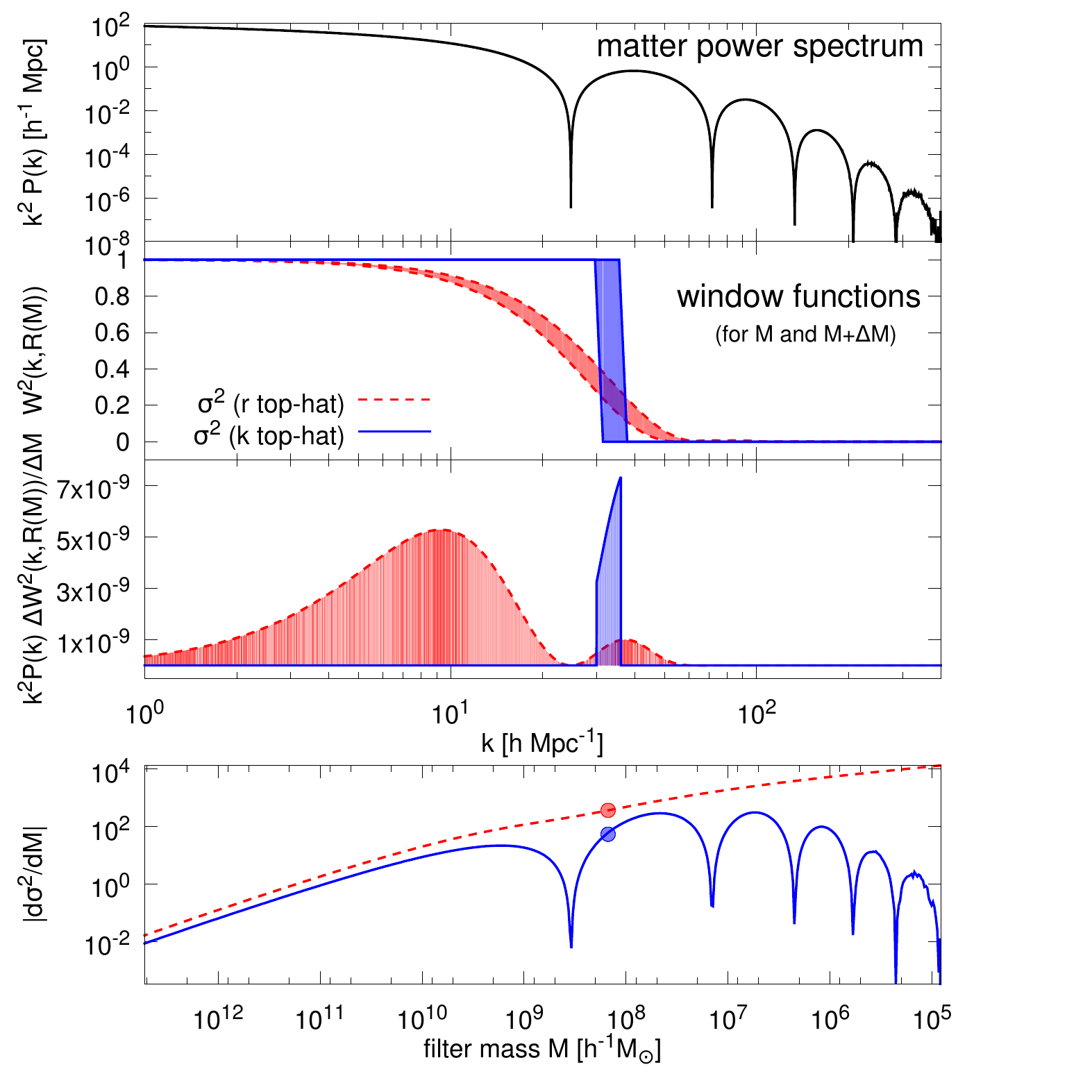}
\caption{Real-space and $k$-space top-hat window functions in Press-Schechter HMF predictions for $\gamma$CDM. The upper panel shows the matter power spectrum, while the second panel shows the Fourier transform of the two window functions ($r$ top-hat and $k$ top-hat). Each window function is evaluated for two filter masses, $M$ and $M+\Delta M$. The difference between the two filter masses is highlighted by the shaded region in each case. The third panel shows the result of applying this differential filter to the matter distribution. Finally, the lower panel shows the integrated result for both window functions. The red and blue points are the results for the specific filter mass $M$ used in the middle two panels.}
\label{fig:hmf:window}
\end{figure}

The window function, $W(r,R)$, is in general, arbitrary. However, certain choices of window function are advantageous as they allow for both a sensible definition of the smoothed density field and an semi-analytical solution for the Fourier transform. A real-space top-hat, $W(r,R) = \Theta(1-|r/R|)$, has the advantage of a well-defined smoothing scale, $R$, defined in terms of the halo mass, $M(R)$, as
\begin{equation}
R = {\left( \frac {3 M} {4 \pi \bar \rho } \right)}^{1/3}\ . \label{eq:window:R}
\end{equation}
However, recent papers (\citealt{Schneider:2013MNRAS,Benson:2013MNRAS}) have shown that this choice does not reproduce the HMF for cosmologies with a cut-off in the matter power spectrum at small scales. Instead, the predicted HMF continues to increase with decreasing $M$, while the suppression of primordial matter perturbations demands the opposite. The reason for this behaviour is illustrated in Fig.~\ref{fig:hmf:window}, where the Fourier-transformed real-space top-hat and (intermediate) steps of the HMF calculations are shown by red/dashed lines. For this type of window function, one obtains significant contributions from a wide range of unsuppressed larger scales, which dominate the resulting variance and thus, the predicted HMF.

A $k$-space top-hat window function is only sensitive to local changes in the matter distribution in $k$-space and thus reproduces the expected suppression in the halo abundance for damped power spectra (see Fig.~\ref{fig:hmf:window}, blue/solid lines). However, the mass-smoothing scale relation ($M$--$R$) must now be defined without the simple geometrical justification of Eq.~\eqref{eq:window:R}, which was used in the real-space case.

Here we use the definition of~\cite{laceycole}, which defines the cut-off wavenumber, $k_s$, in relation to the mass, $M$, based on the normalization choice
\begin{equation}
k_s = {\left( \frac {M} {6 \pi^2 \bar \rho } \right)}^{-1/3}\ . \label{eq:window:ks}
\end{equation}
This corresponds to a correction factor of $c \equiv R k_s \approx 2.42$ with respect to Eq.~\ref{eq:window:R}, so that the semi-analytical HMF matches numerical simulations at large scales\footnote{Note that \cite{Schneider:2013MNRAS} and~\cite{Benson:2013MNRAS} follow a very similar approach, but with slightly different values for $c$.}.

Alternatively, \citet{Schneider:2012MNRAS} found that while the $r$-space top-hat did not match 
the results of their $N$-body simulations, an additional mass-dependent correction factor,
\begin{equation}
 \frac{n(M)}{n^\text{ST}(M)} = \left( 1+ \frac{M_\text{hm}}{\beta M}\right)^{-\alpha} \ , \label{eq:hmf:schneiderfix}
\end{equation}
could correct for this, where $\alpha$ and $\beta$ are free parameters. 
\citet{Schneider:2012MNRAS} set $\beta=1$ and found a best-fitting value of $\alpha=0.6$.
As discussed in the next section, we find better agreement with our simulation results 
by setting $\beta=2$; we will refer to this version of Eq.~\ref{eq:hmf:schneiderfix} as the {\it modified} Schneider et al. correction. 

%%%%%%%%%%%%%%%%%%%%%%%%%%%%%%%%%%%%%%%%%%%%%%%%%%%%%
\subsection{Simulated Halo Mass Function}
%%%%%%%%%%%%%%%%%%%%%%%%%%%%%%%%%%%%%%%%%%%%%%%%%%%%%

\begin{figure}
\includegraphics[width=0.49\textwidth]{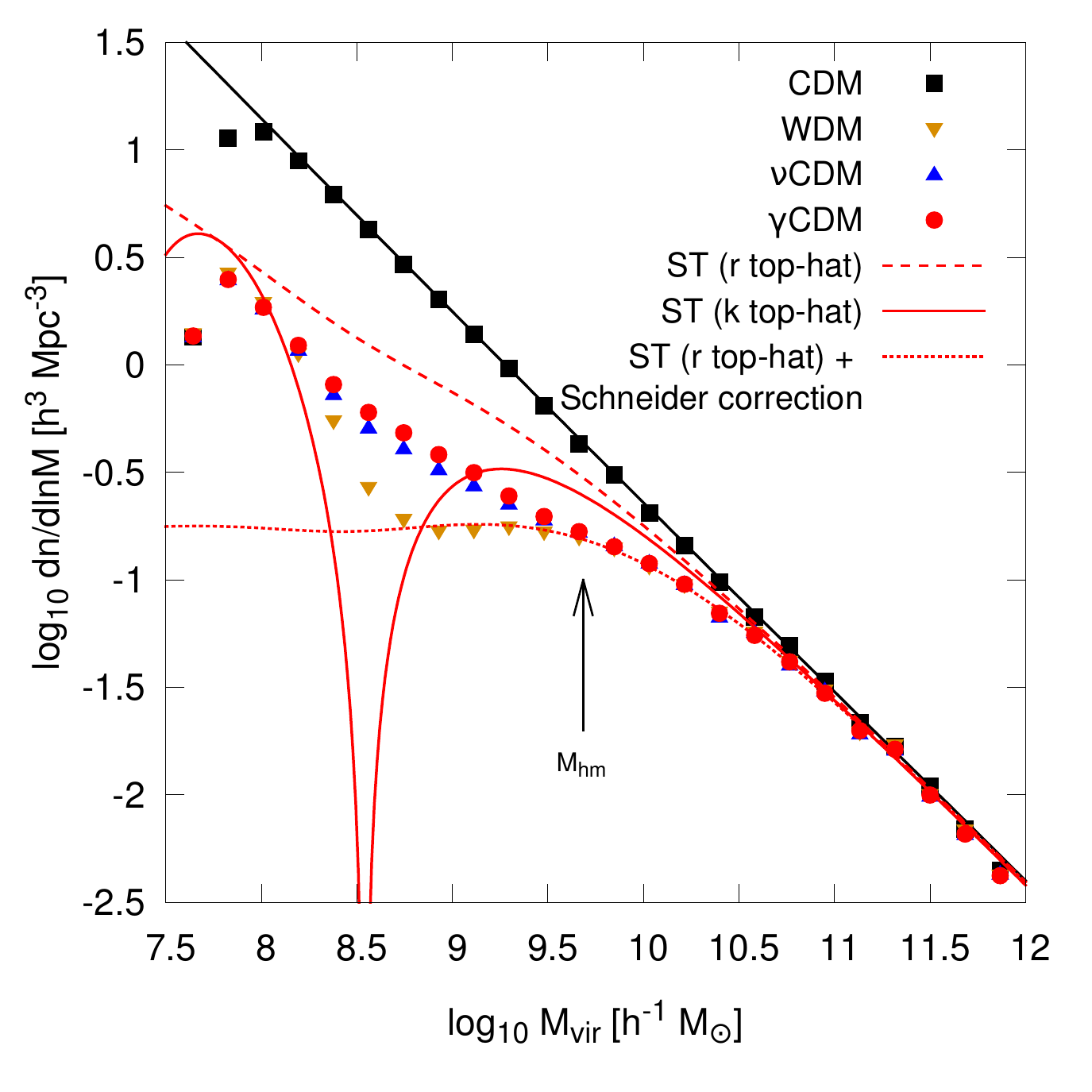}
\caption{The HMFs for collisionless CDM, WDM, $\nu$CDM and $\gamma$CDM at redshift $z = 0$. The HMF measured in models with damped power spectra contains contributions from spurious haloes, which dominate at the smallest masses and result in the upturn seen at $M_{\rm vir} \sim~10^9~h^{-1}~M_\odot$. Above this scale and below the half-mode mass, $M_{\rm hm}$ (marked by the arrow), the abundance of haloes in $\gamma$CDM and $\nu$CDM exceeds that seen in WDM. Predictions using the Sheth-Tormen (ST) formalism with a real-space (dashed) or $k$-space (solid) top-hat window function, as well as the modified Schneider et al. correction (dotted), are also shown. All the semi-analytical predictions fail to predict the HMFs for $\gamma$CDM and $\nu$CDM.}
\label{fig:hmf:hmf}
\end{figure}

We plot the differential HMFs measured in the collisionless CDM, $\gamma$CDM, $\nu$CDM and WDM simulations in Fig.~\ref{fig:hmf:hmf}. We also show the predictions obtained using the semi-analytical approximations described in Sec.~\ref{subsec:psf}.

The mass function proposed by \cite{Schneider:2012MNRAS} predicts fewer haloes than are seen in collisionless CDM but nevertheless overestimates the abundance of haloes less massive than $\sim 10^{10.5}~ h^{-1}~M_{\odot}$. Using a modified version of the Schneider et~al. correction, with $\beta = 2$ instead of $\beta = 1$ extends the reproduction of the simulation results down to a halo mass of $\sim 10^{8.6}~h^{-1}~M_{\odot}$ for WDM. However, it does not reproduce the abundance of haloes seen in the simulations of $\gamma$CDM and $\nu$CDM, underestimating the measured abundance of haloes at $10^{8.6}~h^{-1}~M_{\odot}$ by a factor of two. 

The clear upturn observed in the HMF at low masses in Fig.~\ref{fig:hmf:hmf} (i.e. below $M_{\rm vir} \lesssim~10^{9}~h^{-1}~M_\odot$) is due to non-physical, spurious structures~(\citealt{Wang:2007MNRAS}). We try to avoid contamination from such artificial structures by only considering the mass function and halo properties for objects with masses far above this value.

A comparison between the simulated abundance of haloes in the four models and the semi-analytical predictions reveals significant differences. The main feature, the reduced number of haloes in $\gamma$CDM, $\nu$CDM and WDM, with respect to CDM, is a consequence of the damping of primordial fluctuations on small-scales. There is also a larger number of low-mass structures in $\gamma$CDM and $\nu$CDM, relative to WDM, due to the prominent oscillations in the power spectra of the former models, at wavenumbers larger than the scales on which fluctuations are suppressed.

\begin{figure}
\includegraphics[width=0.48\textwidth]{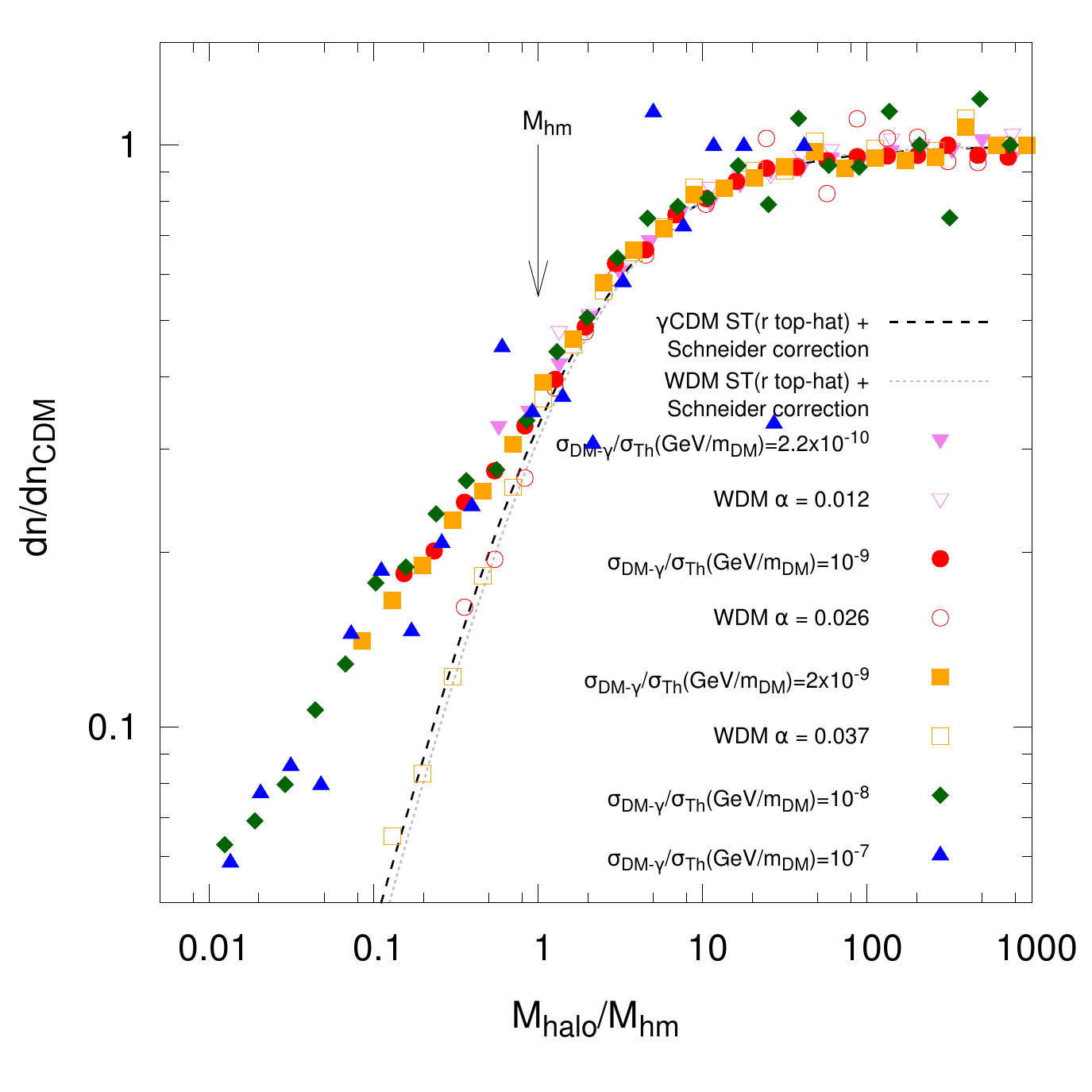}
\caption{The halo abundance expressed in units of the half-mode mass, $M_{\rm hm}$, for $\gamma$CDM (filled symbols) and WDM (unfilled symbols), with respect to CDM, at redshift $z = 0$. The suppression in the HMF is universal with respect to the values of the $\gamma$CDM cross-section and WDM particle mass. The result for WDM matches the semi-analytical prediction of a $r$-space top-hat with the Schneider et al. correction. However, we obtain more haloes in $\gamma$CDM than in WDM as a result of the significant oscillations in the matter power spectrum. Hence, the modified Sheth-Tormen HMF does not provide a good fit to our simulation results.}
\label{fig:hmf:hmf_rel}
\end{figure}

A direct comparison between the $\gamma$CDM and WDM models (see Fig.~\ref{fig:hmf:hmf_rel}) reveals that in both cases, the suppression of the HMF follows a universal profile, if the halo mass is plotted normalized by the half-mode mass, $M_{\rm hm}$. An excess of haloes in $\gamma$CDM with respect to WDM occurs at $M_{\rm hm}$ for all the cross-sections studied in this work. A similar result is found for $\nu$CDM.

The higher halo abundance seen in the $\gamma$CDM and $\nu$CDM simulations compared to that found in WDM is difficult to explain since the primordial matter power spectra shown in Fig.~\ref{fig:theory:ps} are very similar down to the wavenumber corresponding to the half-mode mass, $M_{\rm hm}$. There is a much stronger suppression in the $\gamma$CDM and $\nu$CDM spectra than in WDM immediately below $M_{\rm hm}$. The scales where the power in $\gamma$CDM and $\nu$CDM exceeds that in WDM correspond to halo masses that are an order of magnitude smaller than $M_{\rm hm}$, marked by the location of the first oscillation in the halo abundance for $\gamma$CDM and $\nu$CDM, according to the Sheth-Tormen formalism. Instead of showing a strong reduction in halo abundance below $M_{\rm hm}$, the simulated HMFs for $\gamma$CDM and $\nu$CDM seem to bridge the gap between the primary power cut-off scale and the subsequent increase in the halo abundance resulting from the oscillating matter power spectra. 

Given that the simulations for WDM, $\gamma$CDM and $\nu$CDM use similar initial conditions (e.g. identical box size, phases, number of particles), numerical errors can most likely be excluded as a possible explanation for this deviation. Therefore, this is a strong hint that the understanding of structure formation in the Sheth-Tormen formalism, which works so well in the strictly hierarchical case, appears to fail when there is oscillating power in the initial matter distribution.

%%%%%%%%%%%%%%%%%%%%%%%%%%%%%%%%%%%%%%%%%%%%%%%%%%%%%
\subsection{Halo Bias}
%%%%%%%%%%%%%%%%%%%%%%%%%%%%%%%%%%%%%%%%%%%%%%%%%%%%%

We determine the linear clustering bias of DM haloes, $b_{\rm lin}(M)$, using the ratio between the halo-density cross-correlation and the density-density auto-correlation on large scales (i.e. at small wavenumbers):
\begin{equation}
 b_{\rm lin}(M) = \lim_{k \rightarrow 0} \frac {P_{\rm hm}(M)}{P_{\rm mm}}\ . \label{eq:bias:blin}
\end{equation}
Using the cross-correlation of haloes and mass rather than the autocorrelation of haloes reduces the impact of shot noise (see~\citealt{Angulo:2007ex}).

To ensure that we recover the asymptotic value of $b_{\rm lin}(M)$, we use the largest simulation box of side length 300~$h^{-1}$~Mpc. For large scales ($k \lesssim 0.1~h^{-1}~\mathrm{Mpc}$), convergence is reached as the halo bias becomes constant. Therefore, we can replace the limit in Eq.~\eqref{eq:bias:blin} with the average over all scales larger than $k = 0.1~h^{-1}~\mathrm{Mpc}$ to reduce the impact of statistical fluctuations arising from the small number of high-mass haloes and low-wavenumber modes in the simulation box. This wavenumber scale corresponds to the largest mode in the 100 $h^{-1}$~Mpc box and, as the shot noise fluctuations are less important for the more abundant low-mass DM haloes, we use the smaller box to measure the halo bias for masses below $10^{11}~h^{-1}~M_\odot$.
 
We do not find a significant deviation from the bias expected in collisionless CDM for WDM, $\gamma$CDM or $\nu$CDM, which agrees with the expectations from the semi-analytical models of halo bias. We therefore conclude that the suppression of small-scale structure in the matter power spectra in $\gamma$CDM and $\nu$CDM takes place independently of the linear background in both overdense and underdense regions. Thus, the clustering properties do not change on the mass scales probed here ($M \gtrsim 10^9~h^{-1}~M_{\odot}$).

%%%%%%%%%%%%%%%%%%%%%%%%%%%%%%%%%%%%%%%%%%%%%%%%%%%%%
%%%%%%%%%%%%%%%%%%%%%%%%%%%%%%%%%%%%%%%%%%%%%%%%%%%%%
\section{Results: Halo Properties}
\label{sec:spinconc}
%%%%%%%%%%%%%%%%%%%%%%%%%%%%%%%%%%%%%%%%%%%%%%%%%%%%%
%%%%%%%%%%%%%%%%%%%%%%%%%%%%%%%%%%%%%%%%%%%%%%%%%%%%%

As seen in Sec.~\ref{sec:hmf}, DM--radiation interactions lead to a reduced abundance of low-mass DM haloes. In this section, we focus on three key properties of these haloes: their shape, density profile and spin.

For this analysis, it is important to only consider DM haloes that are dynamically relaxed. We apply the selection criteria presented in \citet{Maccio:2007MNRAS} and \citet{Neto:2007MNRAS}. The DM haloes must satisfy the following conditions\footnote{We omit the substructure mass fraction criterion as this is strongly correlated with the centre-of-mass displacement criterion listed (\citealt{Neto:2007MNRAS}).}:

\begin{itemize}
\item{\it Centre-of-mass displacement}: The offset, $s$, between the halo centre-of-mass, $r_{\rm cm}$, and the potential centre, $r_{\rm cp}$, normalized by the virial radius, $r_{\rm vir}$, satisfies
\begin{equation}
s = \| r_{\rm cp} - r_{\rm cp}\| < 0.07~.
\end{equation}
\item{\it Virial ratio}: The total kinetic energy of the halo particles within $r_{\rm vir}$ in the halo rest frame, $T$, and their gravitational potential energy, $U$, satisfies
\begin{equation}
2T/\|U\| < 1.35~.
\end{equation}
\end{itemize}
These criteria reduce the number of haloes in our sample by a factor of two, but also significantly decrease the scatter as major mergers and their unrelaxed descendants are removed.

In addition to applying these conditions, our mass-averaged results are restricted to: (i) the subset of haloes with a virial mass smaller than $10^{11}~h^{-1}~M_\odot$, i.e. the mass range that shows a suppression in the halo abundance, and (ii) in order to avoid resolution problems, larger than 1000 particles, i.e. mass bins larger than $\sim 10^{9.3}~h^{-1}~M_\odot$. The latter criterion ensures that the estimates for our observables have converged~(\citealt{power:2003MNRAS}). This lower limit also minimizes the possibility of contamination by spurious structures as they form and mainly affect haloes on small mass scales ($M \lesssim 10^9~h^{-1}~M_{\odot}$); this can be checked by studying their contribution to the HMF plotted in Fig.~\ref{fig:hmf:hmf}.

%%%%%%%%%%%%%%%%%%%%%%%%%%%%%%%%%%%%%%%%%%%%%%%%%%%%%
\subsection{Halo Shape}
%%%%%%%%%%%%%%%%%%%%%%%%%%%%%%%%%%%%%%%%%%%%%%%%%%%%%

To characterize the shape of DM haloes, we study the following quantities derived from the three eigenvalues ($a \geq b \geq c$) of the inertia tensor, as calculated by the {\it AMIGA halo finder}:

\begin{itemize}
 \item sphericity: $c/a$
 \item elongation: $b/a$
 \item triaxiality: $ \left(a^2 - b^2\right)/\left(a^2 - c^2\right)$~.
\end{itemize}

\begin{figure}
\begin{center}
\includegraphics[width=0.45\textwidth]{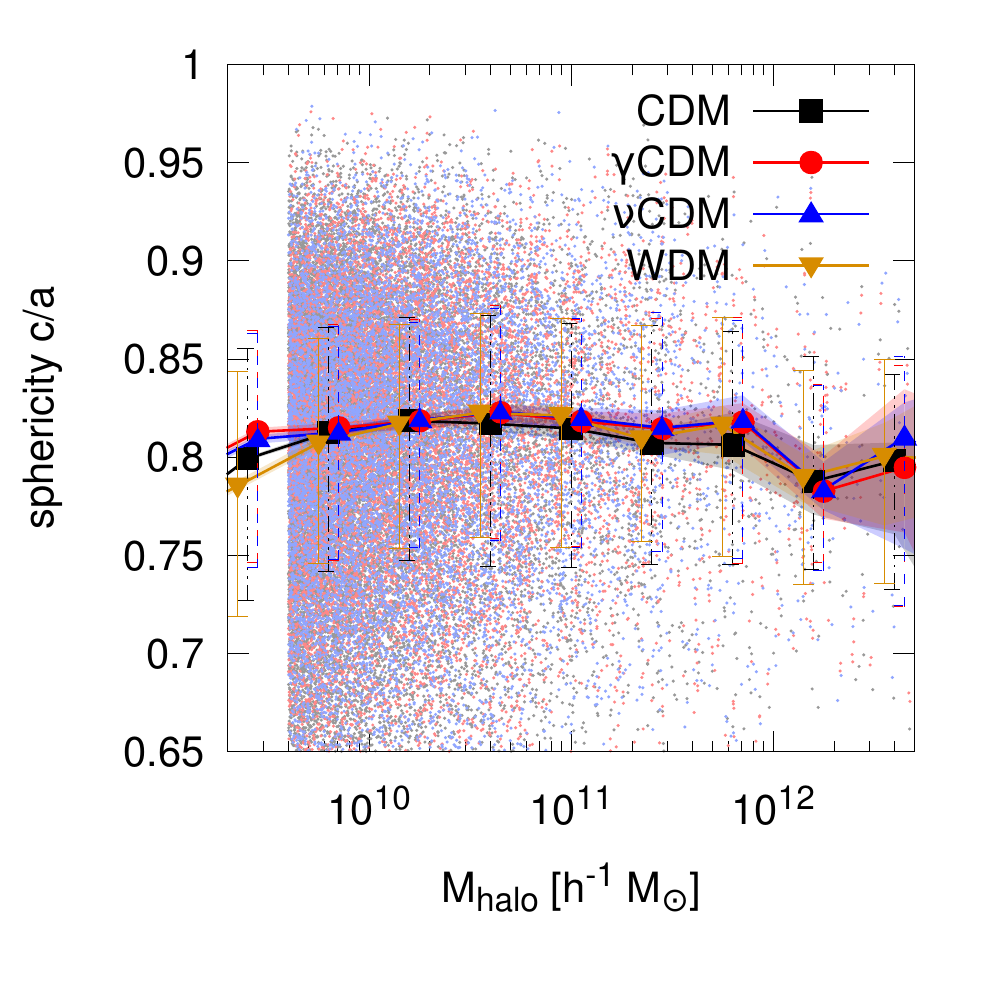}
\vspace{-2ex}
\caption{Sphericity of relaxed DM haloes for CDM, $\gamma$CDM, $\nu$CDM and WDM at redshift $z = 0$. The symbols show the sphericity in mass bins ranging from $4\times10^9~h^{-1}~M_\odot$ to $10^{11}~h^{-1}~M_\odot$ for the different models as labelled. The shaded areas indicate the 95\% CL on the median, given the underlying scatter in the halo sample set (small dots), while the error bars mark the 20\% to 80\% interval for this distribution. The sphericity of DM haloes measured in WDM, $\gamma$CDM and $\nu$CDM shows no significant deviation from CDM.}
\label{fig:spinconc:sphericity}
\end{center}
\end{figure}

\noindent In Fig.~\ref{fig:spinconc:sphericity}, we plot the sphericity measured from the sample set of relaxed haloes. We observe no significant deviation from CDM for WDM, $\gamma$CDM or $\nu$CDM. The same is true for the elongation and triaxiality, and for different redshifts and interaction cross-sections. Thus, we cannot distinguish these models by the shape of their DM haloes.

%%%%%%%%%%%%%%%%%%%%%%%%%%%%%%%%%%%%%%%%%%%%%%%%%%%%%
\subsection{Density Profile and Concentration}
%%%%%%%%%%%%%%%%%%%%%%%%%%%%%%%%%%%%%%%%%%%%%%%%%%%%%

To analyse the density profiles of DM haloes, we first average the density in shells around the centre-of-mass for all haloes in a given mass bin. A comparison of the results with a fitted NFW profile~(\citealt{Navarro:1997ApJ}) reveals a sufficiently good agreement to justify parameterizing the halo profiles in this way\footnote{The fit starts at a minimum radius from the halo centre as defined by \cite{power:2003MNRAS} to ensure convergence of the density profile.}. The NFW profile is completely characterized by the concentration parameter, $c_{\rm NFW}$, which is determined by the halo finder using the approximation presented in~\cite{Prada:2011jf}.

\begin{figure}
\begin{center}
\vspace{-3ex}
\includegraphics[width=0.39\textwidth]{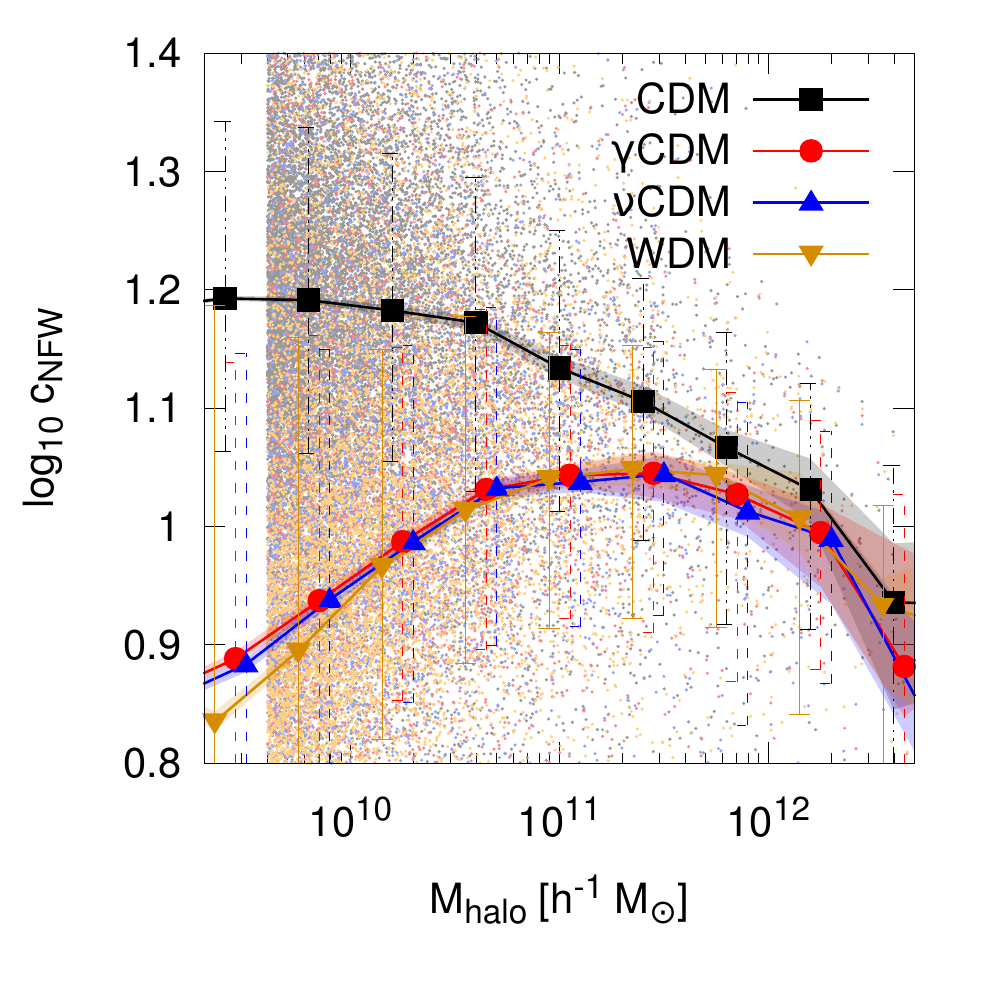}
\vspace{-3ex}
\includegraphics[width=0.39\textwidth]{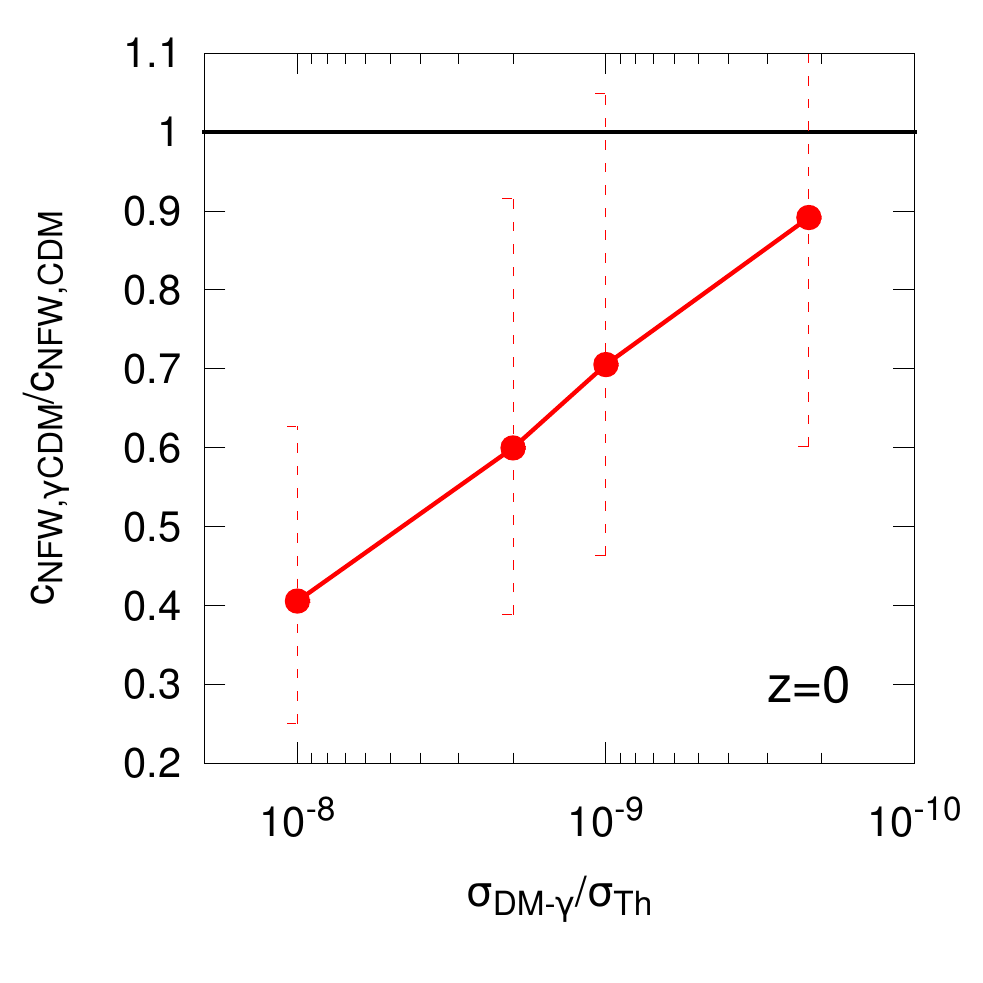}
\vspace{-3ex}
\includegraphics[width=0.39\textwidth]{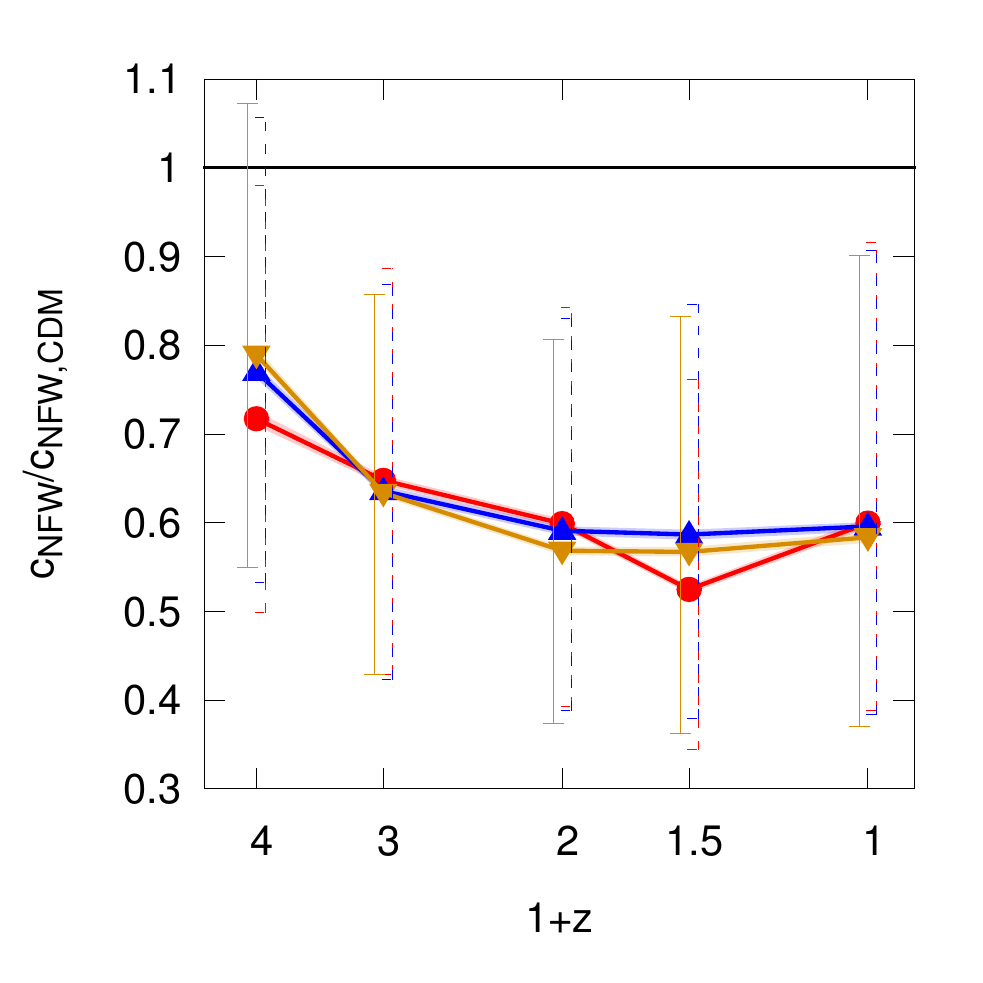}
\caption{The concentration--mass relation (top) shows a strong mass-dependence for $\gamma$CDM, $\nu$CDM and WDM, which develops at scales below $\sim 10^{11}~h^{-1}~M_\odot$. These models are indistinguishable from CDM for more massive haloes. This deviation in the concentration depends strongly on the interaction cross-section (middle) and becomes slightly smaller at higher redshifts (bottom). The data points are the median values for the mass bins ranging from $4 \times 10^9~h^{-1}~M_\odot$ to $10^{11}~h^{-1}~M_\odot$, while the shaded regions mark the 95\% CL on the median, given the underlying scatter in the halo sample set (small dots in the top plot). The error bars mark the 20\% to 80\% interval for this distribution.}
\label{fig:spinconc:conc}
\end{center}
\end{figure}

In Fig.~\ref{fig:spinconc:conc}, we plot the concentration versus mass, cross-section and redshift relations. We observe a significantly lower median value of $c_{\rm NFW}$ in the mass bins below the half-mode mass for $\gamma$CDM and $\nu$CDM compared to CDM. This reduction in concentration with increasing interaction cross-section is similar to the effect seen in WDM simulations with reducing particle mass, which has been explained as being due to the delayed formation time of low-mass haloes~\citep{Lovell:2011rd}. At these late times, the interacting DM models become (effectively) non-collisional for the cross-sections studied here, in the same way that free-streaming in WDM models becomes negligible at low redshifts. Therefore, it is valid to assume that this lower concentration also originates from the later collapse of the DM haloes in these models.

As we increase the interaction cross-section, the deviation from CDM becomes larger due to an increase in the mass scale of the suppression. Since we have fixed the mass interval, the median concentration decreases as a larger number of high-mass haloes become affected.

%%%%%%%%%%%%%%%%%%%%%%%%%%%%%%%%%%%%%%%%%%%%%%%%%%%%%
\subsection{Halo Spin}
%%%%%%%%%%%%%%%%%%%%%%%%%%%%%%%%%%%%%%%%%%%%%%%%%%%%%

We quantify the spin of DM haloes using the ``classical'' definition of \cite{Peebles:1969ApJ}:
\begin{equation}
 \lambda = \frac{J |E|^{1/2}} {G M^{5/2}_{\rm vir}}~,
\end{equation}
where $J$ and $E$ are, respectively, the total angular momentum and binding energy of the material within the virial mass, $M_{\rm vir}$, of a halo.

In the linear and quasi-linear regime, halo spin is described reasonably well using {\it tidal torque theory} (hereafter TTT; \citealt{White:1984ApJ}) and originates from tidal interactions between collapsing haloes. In this framework, the angular momentum of a (proto-)galaxy depends on the mass, but also weakly on the formation time. However, it should be noted that comparisons with numerical simulations have revealed that TTT becomes less applicable as haloes approach turn-around and virialization (\citealt{Porciani:2002MNRAS}). It is still an open question whether haloes acquire significant angular momentum due to mergers with other haloes, as well as from tidal torques~(\citealt{Maller:2002MNRAS,DOnghia:2007MNRAS}).

\begin{figure}
\begin{center}
\vspace{-3ex}
\includegraphics[width=0.39\textwidth]{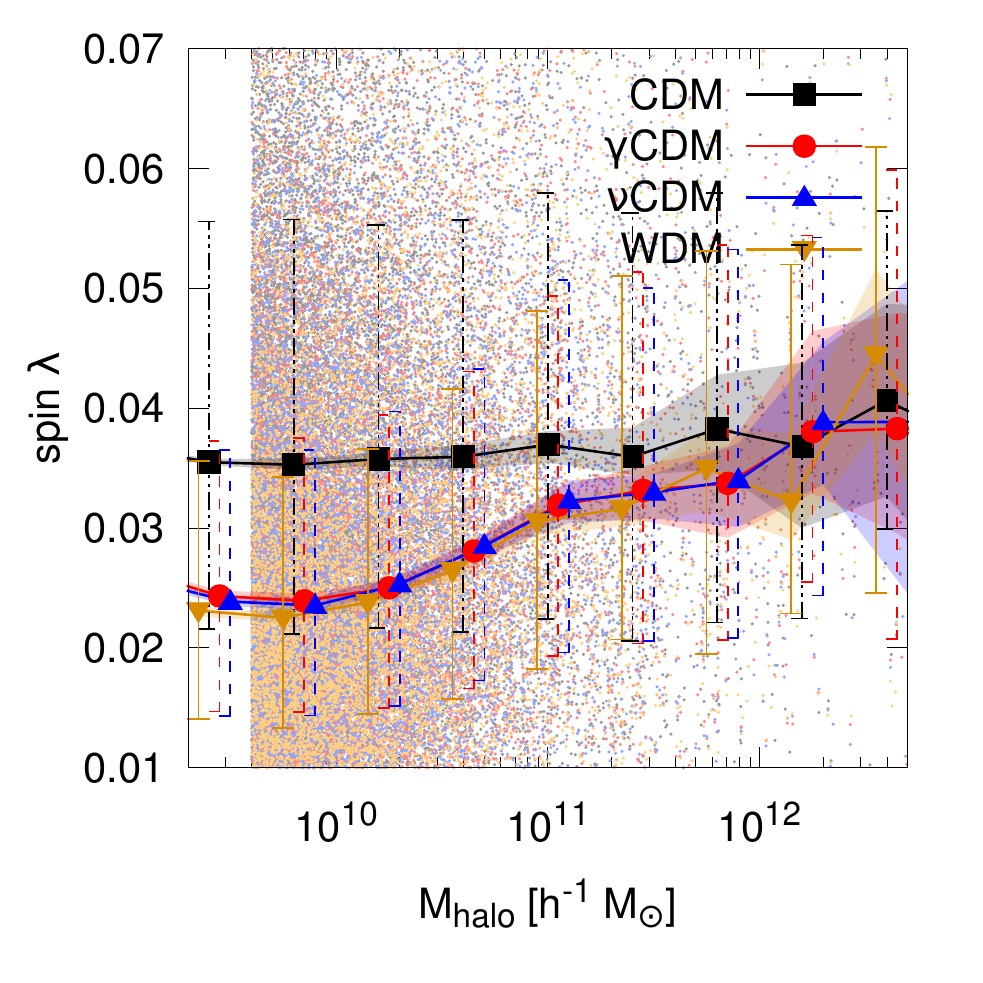}
\vspace{-3ex}
\includegraphics[width=0.39\textwidth]{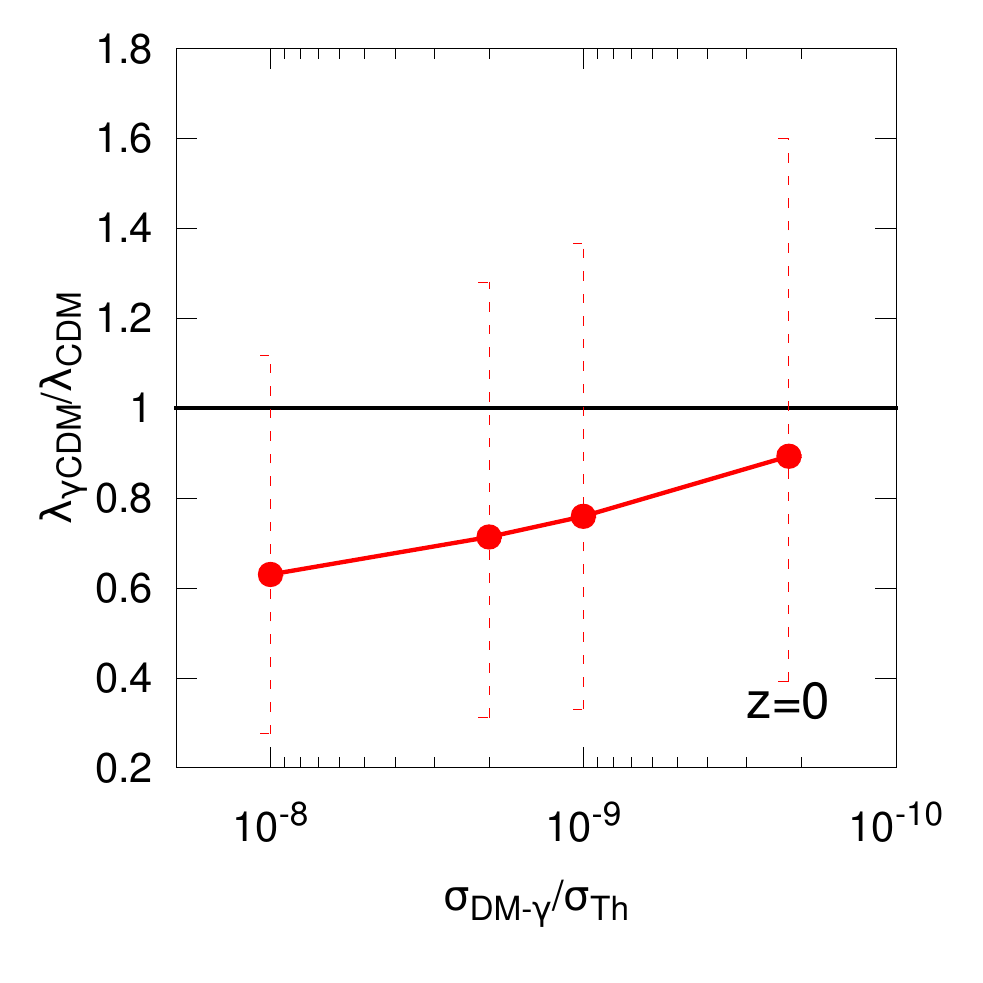}
\vspace{-3ex}
\includegraphics[width=0.39\textwidth]{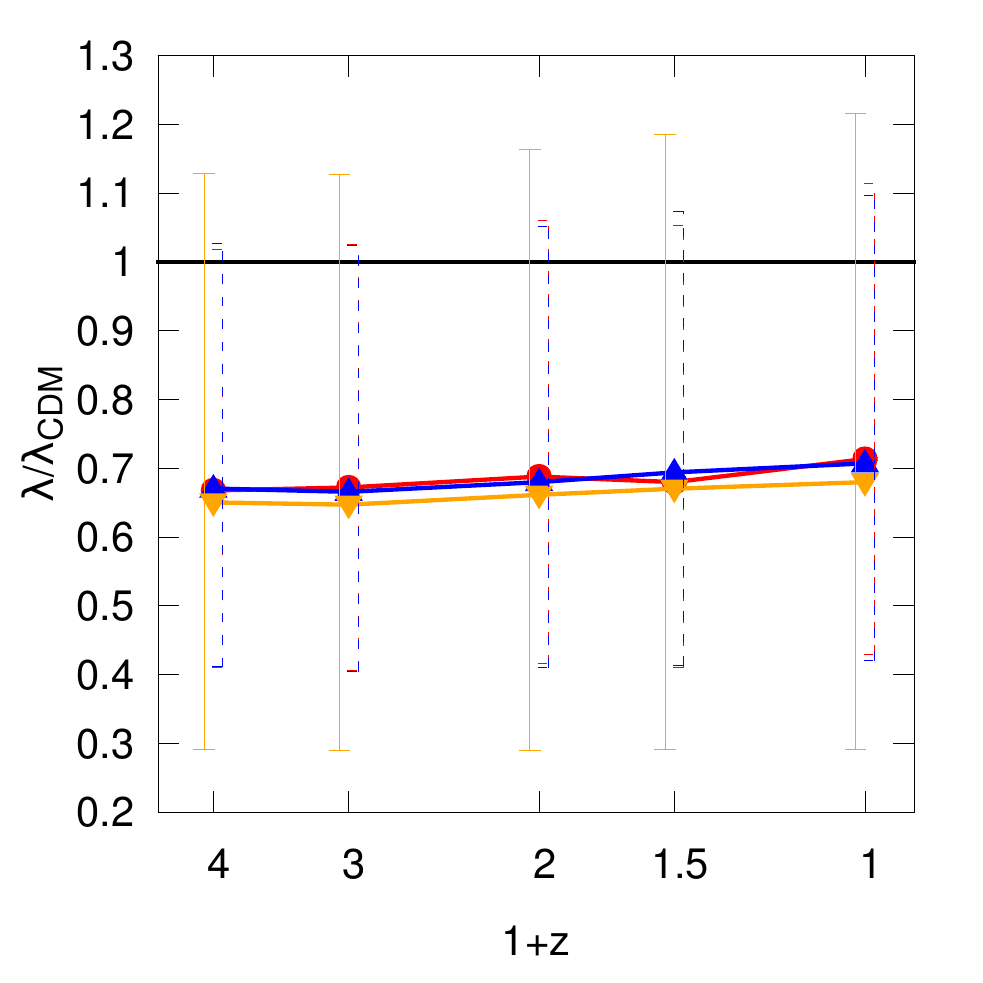}
\caption{The spin--mass relation (top) shows a mass-dependence for $\gamma$CDM, $\nu$CDM and WDM, which develops at scales below $\sim 10^{11}~h^{-1}~M_\odot$. These models are indistinguishable from CDM for more massive haloes. This spin reduction on small scales depends on the interaction cross-section (middle) while the relative deviation from collisionless CDM remains constant over time (bottom). The data points are the median values for the mass bins ranging from $4 \times 10^9~h^{-1}~M_\odot$ to $10^{11}~h^{-1}~M_\odot$, while the shaded regions mark the 95\% CL on the median, given the underlying scatter in the halo sample set (small dots in the top plot). The error bars mark the 20\% to 80\% interval for this distribution.}
\label{fig:spinconc:spin}
\end{center}
\end{figure}

In Fig.~\ref{fig:spinconc:spin}, we plot the median halo spin against virial mass for the different models. We find a similar reduction and evolution of halo spin for $\gamma$CDM, $\nu$CDM and WDM, compared to CDM.

There are various explanations for the difference in halo spin with respect to CDM. As this effect is seen for haloes consisting of more than a few thousand particles, we can rule out a numerical convergence problem. If it originates solely from tidal torques, then the weak dependence of angular momentum on formation time would yield a smaller spin for the earlier formation time found. If mergers are responsible for spinning up haloes, then the lack of smaller progenitors of low-mass haloes and consequently, smoother accretion on to these haloes in $\gamma$CDM, $\nu$CDM and WDM, would also result in a lower net spin. The fact that the difference remains constant over time while the absolute value grows, seems to support the idea that not only the initial tidal torque on the collapsing structure, but also the environment at late times, influences the spin.

\newpage
%%%%%%%%%%%%%%%%%%%%%%%%%%%%%%%%%%%%%%%%%%%%%%%%%%%%%
%%%%%%%%%%%%%%%%%%%%%%%%%%%%%%%%%%%%%%%%%%%%%%%%%%%%%
\section{Conclusion}
\label{sec:conc}
%%%%%%%%%%%%%%%%%%%%%%%%%%%%%%%%%%%%%%%%%%%%%%%%%%%%%
%%%%%%%%%%%%%%%%%%%%%%%%%%%%%%%%%%%%%%%%%%%%%%%%%%%%%

We have shown that even relatively weak DM--radiation interactions can alter structure formation on small cosmic scales. In \cite{boehm:2014MNRAS}, we showed that the number of Milky Way satellites is reduced when DM has primordial interactions with photons ($\gamma$CDM) or neutrinos ($\nu$CDM) and that the resulting number of satellites can be used to place constraints on the interaction cross-section. In this paper, we have extended our previous analysis to study the abundance of DM haloes and their internal properties, namely their shape, density profile and spin. We have also compared different models ($\gamma$CDM, $\nu$CDM and WDM) in which the power spectrum of density fluctuations is suppressed on small scales. 

The halo mass functions measured in our simulations show that the $\gamma$CDM and $\nu$CDM models contain more haloes than WDM around a mass of $10^{9} \hspace{0.5ex} h^{-1} \hspace{0.5ex} M_{\odot}$ for the parameters considered here. This behaviour is not reproduced by various semi-analytical descriptions of the halo mass function. We note that these mass scales are an order of magnitude larger than the scale on which spurious haloes are expected to make a significant contribution~\citep{Wang:2007MNRAS}. The source of this overabundance of haloes with respect to WDM needs to be addressed but could be due to the choice of models for the initial conditions in WDM.

Both the NFW concentration parameter in the density profile and the spin show departures from CDM for low-mass haloes. The halo shape, on the other hand, is independent of the DM model. The lower halo concentration and angular momentum may be due to the delayed formation time of low-mass haloes in $\gamma$CDM and $\nu$CDM and are similar to the trends seen in WDM. However, it should be noted that these halo properties do not provide a means to distinguish between $\gamma$CDM, $\nu$CDM and WDM.

Ideally, the next step in this study would be to include baryonic physics in our simulations, which may have an impact on some of the results reported in our DM-only simulations. \cite{Bryan:2013MNRAS} have shown that efficient gas cooling results in an increased halo spin, while AGN feedback counters this trend. The mass-concentration relation of the haloes is very similar when baryons are included~(\citealt{Schaller:2014}) and the baryons only affect the radial density profile of the inner core within 5\% of the virial radius, producing a contraction. Recent studies also include a possible coupling of DM with dark radiation~\citep{Buckley:2014PhRvD}, which leads to a similar suppression of initial fluctuations as seen in our models and, depending on the cross-section, should give rise to similar results as those discussed in this paper.

%%%%%%%%%%%%%%%%%%%%%%%%%%%%%%%%%%%%%%%%%%%%%%%%%%%%%
%%%%%%%%%%%%%%%%%%%%%%%%%%%%%%%%%%%%%%%%%%%%%%%%%%%%%

\section*{Acknowledgements}

JAS is supported by a Durham University Alumnus Scholarship and RJW is supported by the STFC Quota grant ST/K501979/1. This work was supported by the STFC (grant numbers ST/F001166/1, ST/G000905/1 and ST/L00075X/1) and the European Union FP7 ITN INVISIBLES (Marie Curie Actions, PITN-GA-2011-289442). This work was additionally supported by the European Research Council under ERC Grant ``NuMass'' (FP7-IDEAS-ERC ERC-CG 617143). It made use of the DiRAC Data Centric system at Durham University, operated by the ICC on behalf of the STFC DiRAC HPC Facility (www.dirac.ac.uk). This equipment was funded by BIS National E-infrastructure capital grant ST/K00042X/1, STFC capital grant ST/H008519/1, STFC DiRAC Operations grant ST/K003267/1 and Durham University. DiRAC is part of the National E-Infrastructure. SP also thanks the Spanish MINECO (Centro de excelencia Severo Ochoa Program) under grant SEV-2012-0249.

\bibliographystyle{mn2e}
\bibliography{DM_rad_haloes}

\label{lastpage}

\end{document}